\documentclass[aps,amsfonts,nofootinbib,superscriptaddress,preprint]{revtex4-1}
\usepackage[normalem]{ulem} 
\makeatletter
\usepackage{amsmath}
\usepackage{cleveref}
\usepackage{amssymb}
\usepackage{amsbsy}
\usepackage{bm}
\usepackage{epsfig} 
\usepackage{color}
\usepackage{slashed}
\usepackage{soul}
\usepackage{comment}
\makeatletter
\usepackage{appendix}
\usepackage{epsfig} 
\usepackage{tikz-feynman}

\newcommand{\stkout}[1]{\ifmmode\text{\sout{\ensuremath{#1}}}\else\sout{#1}\fi}

\newcommand{\ee}{\end{equation}}
\newcommand{\bb}{\begin{equation}}
\newcommand{\eqb}{\begin{eqnarray}}

\newcommand{\eqf}{\end{eqnarray}}

\DeclareMathOperator{\Tr}{Tr}
\usepackage[T1]{fontenc}

\begin{document}
\title{ Quintessence and the Higgs Portal in the Carroll limit}
\author{B. Avila}
\email{benjamin.avila@usach.cl}
\affiliation{Departamento de F\'isica, Universidad de Santiago de Chile, Casilla 307, Santiago, Chile}
\author{J. Gamboa}
\email{jorge.gamboa@usach.cl}
\affiliation{Departamento de F\'isica, Universidad de Santiago de Chile, Casilla 307, Santiago, Chile}
\author{R. B. MacKenzie}
\email{richard.mackenzie@umontreal.ca}
\affiliation{ D\'epartement de Physique, Universit\'e de Montr\'eal, Montr\'eal, QC, H3B 3J7, Canada}
\author{F. M\'endez}
\email{fernando.mendez@usach.cl}
\affiliation{Departamento de F\'isica, Universidad de Santiago de Chile, Casilla 307, Santiago, Chile}
\author{M. B. Paranjape}
\email{paranj@lps.umontreal.ca}
\affiliation{ D\'epartement de Physique, Centre de recherche math\'ematiques and the Institut Courtois, Compl\`exe des Sciences, Universit\'e de Montr\'eal, Montr\'eal, Qu\'ebec, Canada, H2V 0B3}
\begin{abstract}
A cosmological model based on two scalar fields is proposed. The first of these, $\varphi$, has mass $\mu$, while the second, $\chi$, is massless. The pair are coupled through a ``Higgs portal''. First, we show how the model reproduces the Friedmann equations if the square of the mass of the $\varphi$ field is proportional to the cosmological constant and $\chi$ represents the quintessence field. Quantum corrections break the conformal symmetry, and the $\chi$ field acquires a mass equal to $\sqrt{3g\Lambda}$.
{{The perturbative approach with $g\ll 1$ is consistent with the bounds for $m_\chi$; moreover, by using dimensional analysis, we estimate
$m_\chi \ll H_0\approx 10^{-33}$ eV, which is in accordance with what is expected in the quintessence scenario.}}
The acceleration of the universe is proportional to 
$\chi^2$, we conclude that for very long times, the solution of the equation of motion approaches $\langle \chi\rangle \sim {m_\chi}/{{\sqrt\lambda}}$ and the universe continues to accelerate, with a constant acceleration.
\end{abstract}
\maketitle

The observation of the accelerated expansion of the universe \cite{riess,perl} is a remarkable result.  Understanding this acceleration will probably require the incorporation of radical new ideas, especially to explain the physics of the late universe. Since its discovery, several approaches have been developed.  Among the most popular are the cosmological constant \cite{weinberg,carroll}, quintessence \cite{wete,ratra1,caldwell}, and phantom energy \cite{caldwell1}, which are consistent with classical tests and current observational data.  However, there is an impression in the community that some aspects of the description are not fully understood.

At the center of current discussions is the idea that the universe is undergoing a period of accelerated expansion similar to inflation \cite{guth1,guth2,linde1,linde2} that could be formally explained by a positive cosmological constant or, more generally, by a fundamental scalar field called quintessence \cite{review}. If we write the Friedmann equations,  quintessence is disguised as a cosmological constant, and the quintessence field provides a dynamical content to dark energy.  Although this idea was proposed in the late eighties in the context of the cosmological constant problem  \cite{wete,ratra1}, it was repurposed as a mechanism to explain the accelerated expansion of the universe that was observed a decade later \cite{caldwell, others}.

Observations show that the universe accelerates predominantly due to the cosmological constant plus contributions attributed to the scalar field. This perspective leads us to think that if the scalar field is fundamental, its mass should be extremely light and much below the characteristic masses of particle physics.  The relevant question is: How does the mass emerge, and how can we infer that it must be very light?

In this article, we analyze the formulation of this cosmological problem from the paradigm of quantum field theory  (QFT) in flat space-time.  At first glance, it may be surprising that the cosmological problem can fit into this paradigm. However, as in other areas of physics \cite{loop}, using the correct variables can sometimes provide insights and simplify otherwise intractable issues; we will argue that this is the case here.

Consider the Lagrangian\footnote{{The sign of the $\varphi$ mass term in (1) has been chosen for reasons that will be clear when comparing with the Friedmann equations (see (\ref{3})-(\ref{4x})).}}
\bb
{\cal L} = \frac{1}{2} (\partial_\mu \varphi)^2 +\frac{1}{2} \mu^2 \varphi^2 +  \frac{1}{2} (\partial_\mu \chi)^2  + \frac{g}{2} \varphi^2 \chi^2 - V(\chi)\label{model1}
\ee
where $\varphi$ and $\chi$ are two scalar fields coupled through what is called the ``Higgs portal'' term proportional to $\varphi^2\chi^2$ and where
\bb
V(\chi) = \frac{\lambda}{4}  \chi^4 \label{pot1}
\ee
is a renormalizable potential.\footnote{The choice of this potential is dictated by the general criterion of renormalizability; the presence of a $\chi^3$ term may have relevance, as in the original \cite{guth1} inflation scenario, if the universe underwent a first-order phase transition. Currently, it is not known if this was the case, so we will exclude the cubic term in our analysis for simplicity.}  As we want to make contact with cosmology, we have to analyze \eqref{model1} in the strong-coupling limit,\footnote{Although the strong-coupling limit has been extensively studied in lattice field theory, this limit and the connection with the Carroll group (the limit $c\to 0$) was largely ignored and only in the last few years has a more systematic analysis been undertaken. For a discussion in the context of the present paper, see \cite{obers,banerjee,nos2}.} that is, when
\bb
\frac{|\partial_t \Phi|}{\Phi} \gg \frac{|\nabla \Phi|}{\Phi}, \label{inequality}
\ee
where $\Phi$ stands for either $\varphi$ and $\chi$.  For the inequality \eqref{inequality} to hold, it is necessary that a typical time interval $\Delta t$  and a typical length scale $\ell$ obey
\bb
\frac{1}{c \Delta t}\gg \frac{1}{\ell}, \label{cineq}
\ee
for both $\varphi$ and $\chi$.  This inequality is satisfied when $c\to 0$, which corresponds to the so-called Carroll contraction of the Poincar\'e group \cite{levy}. The Carroll contraction can also be obtained as the limit $\ell\to \infty$.  It is seen as a way of parameterizing low-energy physics \cite{nos2}.  Then in the strong-coupling limit, the Lagrangian \eqref{model1} can be approximated by dropping the spatial derivatives, giving \cite{martin}
\bb
{\cal L} = \frac{1}{2} {\dot \varphi}^2 + \frac{1}{2} \mu^2 \varphi^2 +  \frac{1}{2} {\dot \chi}^2  + \frac{g}{2} \varphi^2 \chi^2 - V(\chi), \label{model11}
\ee
which is essentially a two-dimensional (classical or quantum) mechanics problem.
For convenience we reparameterize ${\dot \varphi}$ as 
\bb
{\dot \varphi}(t) \to  \frac{ds}{dt}\frac{d\varphi}{ds} \equiv \frac{1}{N}\frac{d\varphi}{ds}.
\ee
where $N= {dt}/{ds}$.  The freedom to choose $s(t)$ arbitrarily introduces a redundancy in the theory, reparameterization invariance, and a consequent gauge degree of freedom which will have to be fixed.  With this reparameterization, \eqref{model11} gives
  \bb
 S= \int dt\,{\cal L} = \int\left(\frac{1}{2N}{\dot \varphi}^2 + \frac{1}{2} \mu^2  N \varphi^2  + \frac{1}{2N}{\dot \chi}^2  + \frac{g}{2} N\,\varphi^2 \chi^2 -N\,V(\chi)\right) ds  \label{10x}
   \ee
where now $\dot\varphi={d\varphi}/{ds}$.  In the gauge $N=1$, the equations of motion are
\eqb
 {\ddot \varphi}  &=& \mu^2  \varphi + g \chi^2 \varphi, \nonumber
 \\
 {\dot \varphi}^2+ {\dot \chi}^2 &=& \mu^2 \varphi^2 + g \varphi^2 \chi^2-2V(\chi), \label{11xxx}
 \\
 {\ddot \chi} &=& g \varphi^2 \chi - V'(\chi),  \nonumber
 \eqf
 where the second equation, obtained by varying with $N$ before setting it equal to 1, simply asserts the conservation of energy.
 
If we make the following judicious non-linear change of variables for the field $\varphi$
  \bb
   \varphi^2= \frac{4}{9} \alpha^2 a^3, \label{5}
   \ee
where $\alpha$ has canonical dimension $+1$, then the equations of motion \eqref{11xxx} become  
    \eqb 
   &&2 \frac{\ddot a}{a} + \left(\frac{{\dot a}}{a}\right)^2 = \frac{4}{3}( \mu^2 + g \chi^2), \label{3}
   \\
   && \left(\frac{\dot a}{a}\right)^2   =\frac{4}{9}(\mu^2 + g \chi^2)- \frac{2}{\alpha^2 a^3}{(\frac{{\dot \chi}^2}{2}+V(\chi))}, \label{3x}
   \\
   && {\ddot \chi} = \frac{2}{3}g ~ a^{3/2} \chi - V'(\chi).
   \label{4x}
   \eqf 
Note that the first two of these are the Friedmann equations with a cosmological constant if we identify $a(s)$ with the scale factor and
   \bb 
   \mu^2 \to  \frac{3}{4} \Lambda. \label{deff1}
   \ee

   To analyze the RHS of \eqref{3}, \eqref{3x}, let us write\footnote{We are using units such that $8  \pi G=1$.}
  \eqb 
   &&2 \frac{\ddot a}{a} + \left(\frac{{\dot a}}{a}\right)^2 = - p, \label{3xx}
   \\
   && \left(\frac{\dot a}{a}\right)^2   =\frac{1}{3} \rho. 
   \label{4xx}
   \eqf 
   and therefore
   \eqb
   -p&=& \Lambda +\frac{4}{3} g \chi^2, \nonumber
   \\
   \rho&=& \Lambda + \frac{4}{3} g \chi^2 -\frac{2}{3\alpha^2 a^3} V(\chi), 
   \eqf
   where we have used the fact that ${\dot \chi} \ll V(\chi)$ \cite{caldwell}.
   The equation of state is 
   \eqb
   -\frac{p}{\rho} &=& 1+   \frac{2}{3 \alpha^2 a^3}\cdot \frac{V(\chi)}{\Lambda + \frac{4}{3} g \chi^2} +\cdots \nonumber
   \\
   &\approx&1 +  {\cal O}(1/a^3), \label{de}
   \eqf
   since in the late universe $a(t) \to \infty$, and \eqref{de} describes dark energy.

%
   
To discuss the new possibilities that could come from quantum theory, let us consider the partition function associated with the action \eqref{10x} (in Euclidean space):
  \bb
  {\cal Z} = {\cal N}\int_0^\infty d N \int {\cal D} \varphi {\cal D} \chi\, e^{-S[\varphi,\chi,N]}. \label{ifunc1}
  \ee
Here 
  \bb
  S[\varphi,\chi,N]= {\cal V} \,\int_{t_1}^{t_2} ds \left( \frac{1}{2N}{\dot \varphi}^2 - \frac{1}{2} \mu^2  N \varphi^2  + \frac{1}{2N}{\dot \chi}^2  - \frac{g}{2} N\,\varphi^2 \chi^2 +N\,V(\chi)\right),
  \ee 
and $ {\cal V}$ is a volume that we will normalize such that ${\cal V}=1$ (we will discuss more about this below).  

  Integrating over $\varphi$, for which the action is quadratic, we encounter the ratio of determinants
  \bb
  J=\left[  \frac{\det \left(-N^{-2}\partial_s^2 -\mu^2 -g\chi^2\right)}{ \det \left(-N^{-2}\partial_s^2 -\mu^2 \right)}\right]^{-\frac{1}{2}}. \label{dett2}
  \ee 
  This ratio can be calculated perturbatively. Let us write
  \eqb
  J&=& \det  \left( 1 + g G \chi^2\right) \nonumber
  \\
  &=& e^{\Tr \ln \left(1 + g G\chi^2\right)}. \label{dett3}
  \eqf
where $G=   \left(-N^{-2}\partial_s^2 -\mu^2\right)^{-1}$ is the propagator. In momentum space the propagator is given by
  \bb
  \tilde G(p_0) = \frac{1}{p_0^2 - \mu^2};
  \label{prop}
  \ee
in position space, it becomes
  \bb
  G(s) = {\cal V}\int \frac{dp_0}{2\pi} \cdot \frac{e^{i p_0 s}}{p_0^2 - \mu^2}, 
  \ee  
  where $s= N(t_2-t_1)=N \Delta t$. Notice that the volume appears again as a multiplicative divergence, but now it can be more clearly seen that it can be absorbed in the renormalization of the theory.  

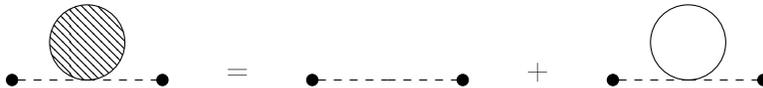
\begin{figure}[h!]
\centering
\begin{tikzpicture}[scale=1,baseline=(b.base)]
\begin{feynman}
\diagram [horizontal=a to b, layered layout] {
  a [dot] -- [scalar] b -- [scalar] c [dot]
};
\path (b)--++(90:0.5) coordinate (A);
\draw [blob] (A) circle(0.5) node[left] at (1.0,0.3) {};
\end{feynman}
\end{tikzpicture} ~~~~ = ~~~~
\begin{tikzpicture}[scale=.8,baseline=(b.base)]
	\begin{feynman}
		\diagram [horizontal=a to b, layered layout] {
			a [dot] -- [scalar] b -- [scalar] c [dot]
		};
	\end{feynman}
\end{tikzpicture}
 ~~~~ + ~~~~
 \begin{tikzpicture}[scale=1,baseline=(b.base)]
 	\begin{feynman}
 		\diagram [horizontal=a to b, layered layout] {
 			a [dot] -- [scalar] b -- [scalar] c [dot]
 		};
 		\path (b)--++(90:0.5) coordinate (A);
 		\draw [solid] (A) circle(0.5) node[left] at (1.0,0.3) {};
 	\end{feynman}
 \end{tikzpicture}

\caption{The first two diagrams for the effective action. The tadpole is the only one that generates mass.
}
\label{fig:1}
\end{figure}

  The Feynman diagrams are shown in Fig. \eqref{fig:1} by expanding $\ln (1-x)$.  The first diagram corresponds to the free case,  while the second diagram is the tadpole contribution and describes a  mass term for $\chi$.  The calculation of the tadpole contribution is simple as the integral over the propagator \eqref{prop} is just  an  arctangent, which gives analytically\footnote{The wrong mass sign is obtained when we return to the Minkowski space, which is a peculiarity of this model.}
  \eqb
  J&=& e^{-i\int_{-\infty}^{\infty} ds~ g\,\mu^2 \chi^2} \nonumber
  \\
  &=& e^{-i\int_{-\infty}^{\infty} ds~\frac{1}{2}\tilde m_\chi^2 \chi^2}
    \eqf
  which is a wrong-sign mass term for $\chi$, with $\tilde m^2_\chi={\frac{3}{2} g \Lambda}$.   {{ Thus, the effective potential to first order in $g$ in Minkowski space is}}
 \bb
 V_{\text{eff}}(\chi) = -\frac{\tilde m_\chi^2}{2}  \chi^2 + \frac{\lambda}{4} \chi^4+ {\cal O}(g^2). \label{effe4}
 \ee
 We see that the $\chi$ field acquires a VEV, $\langle \chi\rangle = \frac{\tilde m_\chi}{\sqrt\lambda}$, {\it i.e.} the quintessence field breaks the $Z_2$ symmetry $\chi\to-\chi$.   The resulting mass for the $\chi$ field is $m_\chi=\sqrt{{{3} g \Lambda}}$. 
  
  \vspace{0.1cm}
\vspace{0.01cm}

Let us summarize the results obtained:
\begin{itemize}
\item  The FRW model coupled to quintessence can be seen as a theory of two scalar fields coupled through a Higgs portal. The first field is a nonlinear realization of the scale factor, and it has a mass $m_{\varphi} = \sqrt{\frac{3}{4}\Lambda}$ where $\Lambda$ is the cosmological constant while the second field 
$\chi$ is dynamically generated.
\item By including quantum corrections, the conformal and $Z_2$ symmetries are broken \cite{jackiw}, and the $\chi$ field acquires mass $m_\chi= \sqrt{3 g \Lambda}$, so that
\[ 
\frac{m_\chi} {m_\varphi} = 2\sqrt{g}\ll 1. 
\]
Thus, the $\chi$ field is extremely light, possibly many orders of magnitude below the mass scales of particle physics.
\end{itemize}

{{We can check the consistency of our result with the bounds of $m_\chi$ found in the literature \cite{tsujikawa} ($m_\chi\lesssim H_0$). Indeed, since $m_\chi^2 \sim  g\Lambda $, for the coupling constant one has $g \sim m_\chi^2/\Lambda$. Therefore
$$
g\lesssim\frac{H_0^2}{\Lambda}\sim10^{-60},
$$
using the observed values $H_0\sim10^{-33}$ eV and $\Lambda\sim 10^{-6}$ eV$^2$. This is consistent with our starting assumption $g\ll 1$.

Parenthetically, we might proceed in analogy with a rough estimate of the fine structure constant in QED where one identifies energy and length scales ${\cal E}$ and $\ell$, respectively. An argument can be used to calculate $\alpha = e^2$ using the hydrogen atom. There, the experimentally measured hydrogen ground state energy $E_0$ and Bohr radius $a_0$ can be used to compute $\alpha$: $|E_0| = e^2/a_0$ giving $\alpha=|E_0| a_0$, which gives the correct value. 

Applying this analysis to the present case is not direct, one should compare the energy in one Hubble volume due to the cosmological constant to the energy it takes to create one Hubble volume full of the condensate of the quintessence field.  However, we can see from dimensional analysis,  if we choose  the length scale  $\ell \sim H_0^{-1}$ (the size of the universe), and the energy scale 
${\cal E} \sim \sqrt{\Lambda}$ (since dark energy constitutes almost 70\% of the energy of the universe)  then, in analogy with QED, one has ${\cal E}\ell = {g}$, or 
$$
g\sim\left(\frac{m_\chi}{H_0}\right)^{2/3}.
$$
Although this argument is not compelling,  it is also not inconsistent.   The limit $g\ll 1$ implies $m_\chi \ll H_0\approx 10^{-33}$ eV, which is well below the mass scales of particle physics, in line with quintessence scenarios, where a tiny mass is expected \cite{turner}.}}

The solutions of the equation of motion with the effective potential \eqref{effe4} are instantons.  However, in QFT, a double-well potential for a scalar field does not permit tunneling due to the infinite space-time volume.  
Surely this suppression of tunneling persists in the Carroll limit.  Since the acceleration of the universe is proportional to $\chi^2$, we conclude that for very long times, the solution of the equation of motion goes to $\langle \chi\rangle \sim {m_\chi}/{\sqrt{\lambda}}$ and the universe continues to accelerate, with a constant acceleration.

{{In this paper, we have taken a different point of view than what is generally assumed \cite{some}. We propose that the mass of the ultralight boson is generated by quantum corrections. As a result of this assumption, we have found a simple effective theory for the scalar $\chi$. This effective theory includes symmetry breaking and a tiny mass that is consistent with other approaches.}}


\section*{Acknowledgements} 
J.G. and F.M would like to dedicate this paper to the memory of their friend and colleague F.~A.~Schaposnik (1947-2023).  This research was supported by DICYT 042131GR and Fondecyt 1221463 (J.G.) and 042231MF (F.M.). We thank NSERC of Canada and the Affaires Internationales Office of the Universit\'e de Montr\'eal for financial support. 
%
%


\begin{thebibliography}{99}
\bibitem{riess}
A.~G.~Riess \textit{et al.} [Supernova Search Team],
Astron. J. \textbf{116} (1998), 1009-1038.
\bibitem{perl} S.~Perlmutter \textit{et al.} [Supernova Cosmology Project],
Astrophys. J. \textbf{517} (1999), 565-586.
\bibitem{weinberg} S.~Weinberg,
Rev. Mod. Phys. \textbf{61} (1989), 1-23.
\bibitem{carroll} S.~M.~Carroll,
Living Rev. Rel. \textbf{4} (2001), 1.

\bibitem{ratra1} B.~Ratra and P.~J.~E.~Peebles,
Phys. Rev. D \textbf{37} (1988), 3406
doi:10.1103/PhysRevD.37.3406.
\bibitem{wete} C.~Wetterich,
Nucl. Phys. B \textbf{302} (1988), 668-696
doi:10.1016/0550-3213(88)90193-9.
\bibitem{caldwell}  R.~R.~Caldwell, R.~Dave and P.~J.~Steinhardt,
Phys. Rev. Lett. \textbf{80} (1998), 1582-1585
doi:10.1103/PhysRevLett.80.1582.
\bibitem{caldwell1} R.~R.~Caldwell,
Phys. Lett. B \textbf{545} (2002), 23-29.

\bibitem{others} A.~Banerjee, H.~Cai, L.~Heisenberg, E.~\'O.~Colg\'ain, M.~M.~Sheikh-Jabbari and T.~Yang,
Phys. Rev. D \textbf{103}, no.8, L081305 (2021)
doi:10.1103/PhysRevD.103.L081305
[arXiv:2006.00244 [astro-ph.CO]].

\bibitem{guth1}
A.~H.~Guth,
Phys. Rev. D \textbf{23}, 347-356 (1981)
doi:10.1103/PhysRevD.23.347.
\bibitem{guth2}
A.~H.~Guth and E.~J.~Weinberg,
Nucl. Phys. B \textbf{212}, 321-364 (1983)
doi:10.1016/0550-3213(83)90307-3.
\bibitem{linde1}
A.~D.~Linde,
Nucl. Phys. B \textbf{216}, 421 (1983)
[erratum: Nucl. Phys. B \textbf{223}, 544 (1983)]
doi:10.1016/0550-3213(83)90072-X.
\bibitem{linde2}
A.~D.~Linde,
Rept. Prog. Phys. \textbf{42}, 389 (1979)
doi:10.1088/0034-4885/42/3/001.



\bibitem{review} For a review, see, for example, 
S.~Tsujikawa,
Class. Quant. Grav. \textbf{30} (2013), 214003.
\bibitem{loop} 
C.~Rovelli,
Living Rev. Rel. \textbf{1}, 1 (1998)
doi:10.12942/lrr-1998-1
[arXiv:gr-qc/9710008 [gr-qc]].
\bibitem{obers} J.~de Boer, J.~Hartong, N.~A.~Obers, W.~Sybesma and S.~Vandoren,
Front. in Phys. \textbf{10}, 810405 (2022)
doi:10.3389/fphy.2022.810405.
\bibitem{levy} J. -M. Lev\'y-Leblond, Annales de l'I. H. Poincar\'e section A, tome 3, no 1 (1965), p. 1-12.

\bibitem{banerjee}K.~Banerjee, R.~Basu, B.~Krishnan, S.~Maulik, A.~Mehra and A.~Ray,
Phys. Rev. D \textbf{108}, no.8, 085022 (2023)
doi:10.1103/PhysRevD.108.085022
[arXiv:2307.03901 [hep-th]].
\bibitem{nos2} J.~Gamboa, J.~Lopez-Sarrion, F.~Mendez and N.~Tapia-Arellano,
[arXiv:2212.02938 [hep-th]]; J.~Gamboa and J.~Lopez-Sarrion,
Int. J. Mod. Phys. A \textbf{36} (2021) no.13, 13
doi:10.1142/S0217751X21500743.

\bibitem{martin} J.~Martin,
Comptes Rendus Physique \textbf{13} (2012), 566-665
doi:10.1016/j.crhy.2012.04.008
[arXiv:1205.3365 [astro-ph.CO]].

\bibitem{jackiw} S.~R.~Coleman and R.~Jackiw,
Annals Phys. \textbf{67} (1971), 552-598
doi:10.1016/0003-4916(71)90153-9.


\bibitem{tsujikawa}
S.~ Tsujikawa,  Class. Quantum Grav. {\bf 30}, 214003 (2013).

\bibitem{turner} 
J.~Frieman, M.~Turner and D.~Huterer,
Ann. Rev. Astron. Astrophys. \textbf{46}, 385-432 (2008)
doi:10.1146/annurev.astro.46.060407.145243.
\bibitem{some}  J. A. Frieman, C. T. Hill, A. Stebbins and I. Waga, Phys. Rev. Lett. {\bf 75}, 2077 (1995); Y. Nomura, T. Watari and T. Yanagida, Phys. Lett. B {\bf 48}, 103 (2000); K. Choi, Phys. Rev.  D {\bf  62}, 043509 (2000); J. E. Kim and H. P. Nilles, Phys. Lett.  B{\bf 553}, 1 (2003).

  \end{thebibliography}
\end{document}